\documentclass[11pt]{article}
\usepackage{amsmath,amssymb,amsfonts}
\usepackage[english]{babel}
\makeatletter
\@addtoreset{equation}{section}
\makeatother

\def\nn{\nonumber} \def\bd{\begin{document}} \def\ed{\end{document}}
\def\ds{\documentstyle}
\let\bm=\bibitem
\newcommand{\be}{\begin{equation}}
\newcommand{\ee}{\end{equation}}
\newcommand{\bea}{\setlength\arraycolsep{2pt} \begin{eqnarray}}
\newcommand{\eea}{\end{eqnarray}}
\newcommand{\hoch}[1]{$\, ^{#1}$}
\def\p{\partial}
\title{\large {\bf Off-shell Noether current and conserved charge in Horndeski theory}}
\date{}

\author{Jun-Jin Peng\footnote{pengjjph@163.com}  \\ \\
\small \sl School of Physics and Electronic Science, Guizhou Normal University, \\
\small Guiyang, Guizhou 550001, People's Republic of China \\
\small \sl Institute of Technical Physics, SEEE, Wuhan Textile University,\\
\small Wuhan, Hubei 430073, People's Republic of China 
}

\begin{document}

\maketitle
\vspace{20pt}

\begin{center}
\textbf{Abstract}
\end{center}
We derive the off-shell Noether current and potential in the context of
Horndeski theory, which is the most general scalar-tensor theory with a
Lagrangian containing
derivatives up to second order while yielding at most to second-order equations
of motion in four dimensions. Then the formulation of conserved charges is
proposed on basis of the off-shell Noether potential and the surface term got
from the variation of the Lagrangian. As an application, we calculate the conserved
charges of black holes in a scalar-tensor theory with non-minimal coupling between
derivatives of the scalar field and the Einstein tensor.

%%%%%%%%%%%%%%%%%%%%%%%%%%%%%%%%%%%%%%%%%%%%%%%%%%%%%%%%%%%%%%%%%%%%%%%%%
\voffset=-.90pt
\vspace{40pt}

%%%%%%%%%%%%%%%%%%%%%%%
\section{Introduction} \label{secone}
%%%%%%%%%%%%%%%%%%%%%%%

Horndeski gravity theory, first formulated by Horndeski back in 1974
\cite{Horndeski}, is the most general scalar-tensor theory with a
single scalar degree of freedom that possesses a Lagrangian containing
higher than second order derivatives while yielding second-order equations
of motion for the metric and the scalar field in four dimensions.
In some cases, it recovers general relativity and a wide class of modified
gravity models with a single scalar hair, such as Brans-Dicke theory,
f(R) gravity, $k$-essence \cite{Kessence} and the covariant Galileon
\cite{CovGalileon}. The Horndeski theory has received extensive
attention since Deffayet et al. re-discovered independently it as the theory of
generalized Galileon \cite{DeayetGSZ}, which is equivalent to the
original Horndeski theory in spite of a different formulation \cite{KobaYY}.
Up to now this theory has been extended and developed to investigate
various aspects associated with gravitational theory, ranging from black hole
physics to cosmology. In the context of solutions in the Horndeski theory,
spherically symmetric solutions were investigated in \cite{KimuKY,KayaNT,KaseT}
while rotating solutions were discussed in \cite{MaSilMiBe}. Owing to its
complexity of the theory, it is difficult to find exact solutions in the
full theory. Consequently, a lot of attentions were drawn to seeking solutions
\cite{FengLLP,CharmT,MomHGRF,KobaTana,Minami,AnabCO,BabiCh,Rinaldi,BabiCH,Cistern,
KolyKPS,BTZBraH,Giribet,MaSilMiBe,CisCDS}
in a particular case of the Horndeski theory, where the Lagrangian only involves
non-minimal coupling between derivatives of the scalar field and the Einstein
tensor \cite{KobaTana,Sushkov}. To further interpret thermodynamic property of
the black hole solutions, it is of great necessity to find a proper formulation
of conserved charges in the full Horndeski theory, which is just our
motivation in this work.

Recently, by making use of the Noether procedure, Kim, Kulkarni and Yi
put forward a formulation of quasi-local conserved charges in covariant
theories of pure gravity \cite{KimKY}, which can be seen as an off-shell
generalization of the conventional Abbott-Deser-Tekin (ADT) formalism
that is defined in terms of the Noether potential got through the
linearized perturbation for the expression of the gravitational field
equation in a fixed background metric satisfying the equation of motion
in vacuum \cite{AbbottD,DeserT}. The main ideas of their method go as
follows. Starting with the variation of the Lagrangian for the gravity
system along the line of the covariant phase space approach
\cite{IyerWald,IyerWald2}, one reads off the expression
for the equation of motion and surface term. Next, under the assumption that the
variation is induced by a diffeomorphism symmetry generated by a smooth
vector field, an off-shell Noether current and its corresponding potential
with respect to the vector field is introduced in terms of the
expression of the field equation and the surface term. Finally, by establishing
the one-to-one relationship between the off-shell Noether potential and
the ADT potential and following the method in \cite{BarnichB,Barnich,BCintegC}
to incorporate a single parameter path in the space of solutions
into the formalism, one can propose a quasi-local formulation of the conserved
charges in the theories of gravity.

Inspired with the generalized formalism for the quasi-local conserved charges
proposed in the work \cite{KimKY}, the method therein has been generalized and
developed to study conserved charges in various gravitational theories
coupled with matter fields or not. In \cite{CiteHJPY}, Hyun, Jeong, Park
and Yi developed the formalism in \cite{KimKY} by considering all the effect
from gravitational field and matter fields in covariant theories of gravity,
and they further showed that conserved charges via the modified ADT formalism
coincide with those by the covariant phase space approach \cite{IyerWald,IyerWald2}.
In \cite{JJPeng}, by directly varying the Bianchi identity for the expression of
the equation of motion, we presented an off-shell Noether current in a different,
but equivalent formulation compared with the one in \cite{KimKY} . Then we
employed the generalized formulation to calculate the quasi-local conserved
charges of black holes in four-dimensional conformal Weyl gravity and in arbitrary
dimensional Einstein-Gauss-Bonnet gravity. Conserved charges of black holes with
a sole scalar hair were taken into account in \cite{AhnHPY,HyunJPY,Wuli}. Other
applications and developments of the modified ADT formalism can be found in
\cite{CiteChernS,MBGhass,ABBGCHJ,Setare,HyunJP,CiteLifBH}.

In this letter, we focus on providing a systematic approach to calculate
the conserved charge in the full Horndeski gravity theory. To do this, we
derive the off-shell Noether currents and potentials of this theory and
follow the works \cite{KimKY,CiteHJPY} to propose a formulation of the
conserved charge through building the one-to-one correspondence between the
off-shell Noether potential and the ADT potential. Then the generalized
formalism is extended to a special scalar-tensor theory with non-minimal
coupling to gravity \cite{KobaTana,Sushkov}, which has attracted much
attention in the context of black hole physics. As concrete examples,
we explicitly compute the mass and angular momentum of the three-dimensional
rotating black hole with a sole scalar degree of freedom in \cite{BTZBraH}
and the mass of the four-dimensional (charged) spherically symmetric
black holes in \cite{AnabCO,Cistern}.

The outline of this letter goes as follows. In section \ref{sectwo}, we derive the
off-shell Noether currents and potentials of the Horndeski theory and then present
the formulation of conserved charge in this theory. In section \ref{secthree}, as
an application in a particular case of the Horndeski theory, the off-shell Noether
potential of the scalar-tensor theory with a single scalar field non-minimally
coupled to the metric is derived. By using the formulation defined in terms of the
potential and the surface term, we compute the mass and angular momentum of
three-dimensional rotating black holes and the mass of four-dimensional charged
spherically symmetric black holes in the special theory. The last section is
our conclusions.

%%%%%%%%%%%%%%%%%%%%%%%%%%%%%%%%%%%%%%%%%%%%%%%%%%%%
\section{Off-shell Noether corrents and the formulation of conserved charges}\label{sectwo}
%%%%%%%%%%%%%%%%%%%%%%%%%%%%%%%%%%%%%%%%%%%%%%%%%%%%

In this section, we derive the off-shell Noether currents and their corresponding
potentials in the framework of the Horndeski theory along the line of the works
\cite{KimKY,CiteHJPY}. By building the relationship between the off-shell Noether
potential and the ADT potential \cite{AbbottD,DeserT}, we further give the
formulation of the conserved charge in the Horndeski gravity theory.

As a starting point, we consider the Lagrangian for the Horndeski theory that takes
the form
\be
L=\sum_{i=0}^3L_{(i)} \, , \label{LagranHorn}
\ee
where the components $L_{(i)}$ are given by \cite{Horndeski,DeayetGSZ,KobaYY}
\bea
L_{(0)}&=&\sqrt{-g} G^{(0)}(\phi,X) \, , \nn \\
L_{(1)}&=&\sqrt{-g} G^{(1)}(\phi,X) (\nabla_\mu \nabla^\mu \phi) \, , \nn \\
L_{(2)}&=&\sqrt{-g} \big[2G^{(2)}_{,X}(\phi,X)\delta_{\nu_1}^{[\mu_1}
\delta_{\nu_2}^{\mu_2]}(\nabla_{\mu_1} \nabla^{\nu_1} \phi)
(\nabla_{\mu_2} \nabla^{\nu_2} \phi)+ RG^{(2)}(\phi,X) \big] \, , \nn \\
L_{(3)}&=&\sqrt{-g} \big[6G^{(3)}_{,X}(\phi,X)
\delta_{\nu_1}^{[\mu_1}\delta_{\nu_2}^{\mu_2}\delta_{\nu_3}^{\mu_3]}
(\nabla_{\mu_1} \nabla^{\nu_1} \phi)(\nabla_{\mu_2} \nabla^{\nu_2} \phi)
(\nabla_{\mu_3} \nabla^{\nu_3} \phi) \, \nn \\
&&-6G_{\mu\nu}(\nabla^\mu \nabla^\nu \phi)G^{(3)}(\phi,X) \big] \, .
\label{ComLagranHorn}
\eea
In the above equation, the scalar curvature $R$ is defined as the trace
of the Ricci tensor $R_{\mu\nu}$, $G_{\mu\nu}$ denotes the Einstein
tensor, $X=-1/2\nabla_\mu\phi \nabla^\mu \phi$,
$G^{(i)}_{,X}(\phi,X)= \partial G^{(i)}(\phi,X)/\partial X$,
and $G^{(i)}(\phi,X)$ are arbitrary functions of the scalar field
$\phi$ and its kinetic term $X$. The
components $L_{(2,3)}$ contain terms involving no-minimal couplings to
gravity, which result in the elimination of higher derivatives that
might appear in the equations of motion. Consequently, in spite of the
Lagrangian containing higher-order derivative terms, the field equations
are of second-order \cite{GaoSteer}. The Horndeski gravity theory is a
general scalar-tensor theory. It includes general relativity and all
popular modified gravity theories with a single scalar field as special cases,
for instance, the Lagrangian (\ref{LagranHorn}) reduces to the conventional
Einstein-Hilbert Lagrangian when $G^{(2)}=1/2$ with $G^{(0,1,3)}=0$, and
in the work \cite{MaSilMiBe}, it has been shown that the Horndeski theory
recovers several well-known modified gravity models, such as Brans-Dicke theory,
f(R) gravity, $k$-essence \cite{Kessence}, the covariant Galileon
\cite{CovGalileon} and the theory of Einstein-dilaton-Gauss-Bonnet gravity,
with proper choice of the four free functions $G^{(i)}$.

The variation of the Lagrangian (\ref{LagranHorn}) is read off as
\bea
\delta L&=&\sum_{i=0}^3\delta L_{(i)}
=\sqrt{-g}\big[\mathcal{T}_{\mu\nu}\delta g^{\mu\nu}+\mathcal{E}_{(\phi)}\delta\phi
+\nabla_\mu\Theta^\mu(\delta g,\delta\phi)\big]         \, , \nn \\
\mathcal{T}_{\mu\nu}&=&\sum_{i=0}^3\mathcal{T}_{\mu\nu}^{(i)} \, , \quad
\mathcal{E}_{(\phi)}=\sum_{i=0}^3\mathcal{E}_{(\phi)}^{(i)} \, , \quad
\Theta^\mu(\delta g,\delta\phi)=\sum_{i=0}^3\Theta_{(i)}^\mu(\delta g,\delta\phi)
\, , \label{VariationLag}
\eea
where and in what follows, the quantity with the index ``$^{(i)}$" is the one
corresponding to the Lagrangian $L_{(i)}$. In Eq. (\ref{VariationLag}), the
expressions of the field equation $\mathcal{T}_{\mu\nu}^{(0)}$,
$\mathcal{T}_{\mu\nu}^{(1)}$, $\mathcal{E}_{(\phi)}^{(0)}$ and
$\mathcal{E}_{(\phi)}^{(1)}$ are given by
\bea
\mathcal{T}_{\mu\nu}^{(0)}&=& -\frac{1}{2}\big(G^{(0)}g_{\mu\nu}
+G^{(0)}_{,X}\Phi_{\mu\nu}\big) \, , \nn \\
\mathcal{T}_{\mu\nu}^{(1)}&=& \frac{1}{2}g_{\mu\nu}\nabla_\sigma G^{(1)}
\nabla^\sigma \phi -\nabla_{(\mu} G^{(1)}\nabla_{\nu)} \phi
-\frac{1}{2}G^{(1)}_{,X}\Box\phi\Phi_{\mu\nu}\, , \nn \\
\mathcal{E}_{(\phi)}^{(0)}&=& \nabla_\mu\big(G^{(0)}_{,X}\nabla^\mu\phi\big)
+G^{(0)}_{,\phi} \, , \nn \\
\mathcal{E}_{(\phi)}^{(1)}&=& \nabla_\mu\big(G^{(1)}_{,X}\Box\phi\nabla^\mu\phi\big)
+G^{(1)}_{,\phi}\Box\phi+\Box G^{(1)} \, . \label{T01Ephi01}
\eea
Here and in the remainder of this work, the symmetric tensor $\Phi_{\mu\nu}$
is defined through $\Phi_{\mu\nu}=(\nabla_\mu\phi) (\nabla_\nu\phi)$.
The surface terms $\Theta_{(0)}^\mu(\delta\phi)$ and
$\Theta_{(1)}^\mu(\delta g,\delta\phi)$ have the forms
\bea
\Theta_{(0)}^\mu&=& -G^{(0)}_{,X}\nabla^\mu\phi\delta\phi \, , \nn \\
\Theta_{(1)}^\mu&=&
\frac{1}{2}G^{(1)}\big(h \nabla^\mu\phi
-2h^{\mu\nu}\nabla_\nu\phi
+2\nabla^\mu\delta\phi \big)
-\delta\phi G^{(1)}_{,X}(\Box\phi)\nabla^\mu\phi
\nn \\
&&-\delta\phi \nabla^\mu G^{(1)}
\, ,\label{SurfT01}
\eea
where and in what follows
\be
h_{\mu\nu}=\delta g_{\mu\nu} \, , \quad
h^{\mu\nu}=g^{\mu\rho}g^{\nu\sigma}h_{\rho\sigma}=-\delta g^{\mu\nu}
\, , \quad
h=g^{\rho\sigma}\delta g_{\rho\sigma} \, . \nn
\ee
The expressions of the field equations and surface terms associated with the Lagrangian
components $L_{(2,3)}$ are much more involved, so we present them in the appendix
\ref{appA}. We have proved that all the expressions for the equations of motion satisfy
\be
2\nabla^\mu\mathcal{T}_{\mu\nu}^{(i)}+\mathcal{E}_{(\phi)}^{(i)}\nabla_\nu\phi
=0 \, , \label{TEphiRelation}
\ee
which result from the constraint that the Horndeski theory has to reserve
diffeomorphism symmetry and directly
lead to that
\be
2\nabla^\mu\mathcal{T}_{\mu\nu}+\mathcal{E}_{(\phi)}\nabla_\nu\phi=0
\, . \label{GenBianchi}
\ee
In the absence of the scalar field, Eq. (\ref{GenBianchi}) reduces to the
usual Bianchi identity $\nabla^\mu G_{\mu\nu}=0$ in general relativity.
In this sense, Eq. (\ref{GenBianchi}) can be treated as a generalized
Bianchi identity for the Horndeski theory.

Let us now consider the variation induced by a diffeomorphism generated by a smooth vector field
$\zeta^\mu$. In other words, the variation of the field $g_{\mu\nu}$ and $\phi$ in
Eq. (\ref{VariationLag}) is replaced by their Lie derivative with respect to the vector
$\zeta^\mu$. With help of Eqs. (\ref{TEphiRelation}) and (\ref{GenBianchi}), we obtain an
off-shell Noether current $J^\mu$, which is read off as
\bea
J^\mu&=&\frac{L}{\sqrt{-g}}\zeta^\mu+2\mathcal{T}^{\mu\nu}\zeta_\nu
-\Theta^\mu(\mathcal{L}_\zeta g,\mathcal{L}_\zeta\phi)
=\sum_{i=0}^3J_{(i)}^\mu\, , \nn \\
J_{(i)}^\mu&=&\frac{L_{(i)}}{\sqrt{-g}}\zeta^\mu +2\mathcal{T}_{(i)}^{\mu\nu}\zeta_\nu
-\Theta_{(i)}^\mu(\mathcal{L}_\zeta g,\mathcal{L}_\zeta\phi)
\, . \label{OffShNoeCurr}
\eea
The off-shell Noether potential $K^{\mu\nu}$, defined through
$J^\mu=\nabla_\nu K^{\mu\nu}$, takes the form
\be
K^{\mu\nu}= \sum_{i=0}^3K_{(i)}^{\mu\nu} \, , \label{OffSheNoPoten}
\ee
where the off-shell Noether potentials $K_{(i)}^{\mu\nu}$, which are associated with
the Noether currents $J_{(i)}^\mu$ through the relations
$J_{(i)}^\mu=\nabla_\nu K_{(i)}^{\mu\nu}$, are presented by
\bea
K_{(0)}^{\mu\nu}&=& 0\, , \nn \\
K_{(1)}^{\mu\nu}&=&2G^{(1)}\zeta^{[\mu}\nabla^{\nu]}\phi \, , \nn \\
K_{(2)}^{\mu\nu}&=&4G^{(2)}_{,X}\big(\Box\phi\zeta^{[\mu}\nabla^{\nu]}\phi
-\zeta_\sigma\Psi^{\sigma[\mu}\nabla^{\nu]}\phi\big)
+4\zeta^{[\mu}\nabla^{\nu]}G^{(2)} +2G^{(2)}\nabla^{[\mu}\zeta^{\nu]}
\, , \label{OffShePote012}
\eea
and
\bea
K_{(3)}^{\mu\nu}&=&6G^{(3)}_{,X}\big[\big((\Box\phi)^2
-\Psi_{\alpha\beta}\Psi^{\alpha\beta}\big)
\zeta^{[\mu}\nabla^{\nu]}\phi
+2(\zeta^\rho\Psi_{\rho\sigma}-\Box\phi\zeta_\sigma)\Psi^{\sigma[\mu}\nabla^{\nu]}\phi\big]
\nn \\
&&-6\big[2\zeta^{[\mu}\nabla_\sigma\big(\Psi^{\nu]\sigma}G^{(3)}\big)
-2\zeta_\sigma\nabla^{[\mu}\big(\Psi^{\nu]\sigma}G^{(3)}\big)
-2\zeta^{[\mu}\nabla^{\nu]}\big(G^{(3)}\Box\phi\big)
\nn \\
&&+G^{(3)}\big(2\zeta_\sigma G^{\sigma[\mu}\nabla^{\nu]}\phi
-2(\nabla_\sigma\zeta^{[\mu})\Psi^{\nu]\sigma}
-\Box\phi\nabla^{[\mu}\zeta^{\nu]}\big)\big]
\, . \label{OffShelPoten3}
\eea
In Eq. (\ref{OffShelPoten3}) and what follows, for brevity, the symmetric tensor
$\Psi_{\mu\nu}$ is defined as $\Psi_{\mu\nu}=\nabla_\mu\nabla_\nu\phi$.

Comparing the off-shell Noether potentials $K_{(i)}^{\mu\nu}$ and $K^{\mu\nu}$ with
the on-shell ones got through Wald's covariant phase space approach \cite{IyerWald,IyerWald2},
one can find that they are equivalent although the Noether currents are different
in both the cases.

Next, assume that the smooth vector field $\zeta^\mu$ respects the symmetry of spacetime,
achieved by a Killing vector $\xi^\mu$. We follow \cite{CiteHJPY} to introduce the off-shell
ADT current $J_{ADT}^\mu$ associated with such a Killing vector by
\bea
J_{ADT}^\mu &=&\delta\mathcal{T}^{\mu\nu}\xi_\nu
+\frac{1}{2}g^{\rho\sigma}\delta g_{\rho\sigma}\mathcal{T}^{\mu\nu}\xi_\nu
+\mathcal{T}^{\mu\nu}\delta g_{\nu\sigma}\xi^\sigma
+\frac{1}{2}\xi^\mu\big(\mathcal{E}_{(\phi)}\delta\phi
+\mathcal{T}_{\rho\sigma}\delta g^{\rho\sigma} \big) \nn \\
&=&\nabla_\nu Q_{ADT}^{\mu\nu}\, , \label{ADTcurr}
\eea
where $Q_{ADT}^{\mu\nu}$ is just the off-shell ADT potential corresponding to
the ADT current. In terms of the variation of the Lagrangian (\ref{LagranHorn}) and
the definition of the off-shell Noether current $J^\mu$, the ADT potential which
is in one-to-one correspondence with the off-shell Noether potential
can be presented by
\bea
Q_{ADT}^{\mu\nu}&=&\frac{1}{2}\frac{1}{\sqrt{-g}}\delta\big(\sqrt{-g}K^{\mu\nu}(\xi)\big)
-\xi^{[\mu}\Theta^{\nu]}(\delta g,\delta\phi)
=\sum_{i=0}^3Q_{(i)}^{\mu\nu}\, , \nn \\
Q_{(i)}^{\mu\nu}&=&\frac{1}{2}\frac{1}{\sqrt{-g}}\delta\big(\sqrt{-g}K_{(i)}^{\mu\nu}(\xi)\big)
-\xi^{[\mu}\Theta_{(i)}^{\nu]}(\delta g,\delta\phi) \, . \label{ADTpotenQ}
\eea
In Eq. (\ref{ADTpotenQ}), the Killing vector $\xi^\mu$ is treated as a fixed background, namely,
$\delta\xi^\mu=0$, and the quantities $Q_{(i)}^{\mu\nu}$ denote the contributions from the
Lagrangian $L_{(i)}$ respectively. For the variation of the off-shell Noether potentials
$K_{(i)}^{\mu\nu}$ , see the equations (\ref{VaOffShePote012}) and (\ref{VaOffShPoten3})
in the appendix \ref{appB}.

Finally, by following the approach in \cite{BarnichB,Barnich,BCintegC} to incorporate
a single parameter path characterized by a parameter $s(s\in[0,1])$ in the space
of solutions, we define the covariant formulation of conserved charges associated with the
Noether potential $Q_{ADT}^{\mu\nu}$ in Eq. (\ref{ADTpotenQ}) by \cite{KimKY,CiteHJPY}
\be
\mathcal{Q}=\frac{1}{8\pi}\int_0^1 ds \int d\Sigma_{\mu\nu} Q_{ADT}^{\mu\nu}(g,\phi;s)
\, , \label{QdefineAn}
\ee
where $d\Sigma_{\mu\nu}=\frac{1}{2}\frac{1}{(D-2)!}
\epsilon_{\mu\nu\mu_1\mu_2\cdot\cdot\cdot\mu_{(D-2)}}dx^{\mu_1}\wedge\cdot\cdot\cdot
\wedge dx^{\mu_{(D-2)}}$ with $\epsilon_{012\cdot\cdot\cdot(D-1)}=\sqrt{-g}$
and $D$ is the dimension of spacetime. Eq. (\ref{QdefineAn}) can be a proposal of
the formalism for
the conserved charge, defined in the interior region or at the asymptotical infinity,
for the most general Horndeski theory with the Lagrangian (\ref{LagranHorn})
whenever its integration is well-defined.

%%%%%%%%%%%%%%%%%%%%%%%%%%%%%%%%%%%%%%%%%%%%%%%%%%%%
\section{Conserved charges in a scalar-tensor theory with non-minimal
derivative coupling}\label{secthree}
%%%%%%%%%%%%%%%%%%%%%%%%%%%%%%%%%%%%%%%%%%%%%%%%%%%%

As an application of the off-shell Noether current and the formulation of the
conserved charge, in the present section, we give a derivation of the
formulation of the conserved charge in the context of a scalar-tensor theory
with the prescription of non-minimal coupling between derivatives of a scalar
field and the Einstein tensor, and then explicitly compute the mass and
angular momentum of (rotating) black holes in such a theory. We start with the
Lagrangian
\cite{KobaTana,Sushkov}
\be
L_{(s)}=\sqrt{-g}[\lambda (R-2\Lambda)-\eta\nabla_\mu\phi \nabla^\mu \phi
+\beta G_{\mu\nu}\nabla^\mu\phi \nabla^\nu \phi]
\, , \label{WAdSLagran}
\ee
where $(\lambda,\Lambda,\eta,\beta)$ are constants. In fact, the Lagrangian
(\ref{WAdSLagran}) can be seen as a subclass of the Lagrangian
(\ref{LagranHorn}) for the full Horndeski theory in addition to a total
divergence term, by setting
\be
G^{(0)}=-2\lambda\Lambda+2\eta X \, , \quad
G^{(1)}=0 \, , \quad
G^{(2)}= \lambda \, , \quad
G^{(3)}=\frac{\beta}{6}\phi \, ,
\ee
or \cite{KobaTana}
\be
G^{(0)}=-2\lambda\Lambda+2\eta X \, , \quad
G^{(2)}= \lambda +\beta X \, , \quad
G^{(1)}=G^{(3)}=0 \, .
\ee
As a consequence, we only need to substitute the above $G^{(i)}$ into the expressions
for the equations of motion, surface terms and off-shell Noether potentials associated
to the most general Lagrangian (\ref{LagranHorn}) to get the corresponding quantities
for the Lagrangian (\ref{WAdSLagran}). The expressions of the field equations are read
off as
\bea
T^{(s)}_{\mu\nu}&=& \lambda G_{\mu\nu}-\eta\Phi_{\mu\nu}
+g_{\mu\nu}(\lambda\Lambda-\eta X)
+\frac{\beta}{2}\big[4R^\sigma_{~(\mu}\Phi_{\nu)\sigma}
-2\nabla^\sigma\nabla_{(\mu}\Phi_{\nu)\sigma}
+\Box\Phi_{\mu\nu} \nn \\
&&+g_{\mu\nu}(\nabla^\rho\nabla^{\sigma}\Phi_{\rho\sigma}
-R^{\rho\sigma}\Phi_{\rho\sigma}+2\Box X)
+2X G_{\mu\nu} -2\nabla_\mu\nabla_\nu X -R\Phi_{\mu\nu}\big] \, , \nn \\
\mathcal{E}^{(s)}_{(\phi)}&=& 2\nabla_\mu\big[(\eta g^{\mu\nu}
-\beta G^{\mu\nu})\nabla_\nu\phi\big]
 \, .   \label{TEphiWAdS}
\eea
The surface term $\Theta_{(s)}^\mu(\delta g,\delta\phi)$ for the Lagrangian
(\ref{WAdSLagran}) has the form
\bea
\Theta_{(s)}^\mu&=&-2\eta\nabla^\mu\phi\delta\phi
+2\lambda g^{\rho[\mu}\nabla^{\sigma]}h_{\rho\sigma}
+\frac{\beta}{2}\big[
2\Phi_{\sigma\rho}\nabla^\sigma h^{\rho\mu}
-2h_{\rho\sigma}\nabla^\rho\Phi^{\sigma\mu}
\nn \\
&&-\Phi^{\rho\sigma}\nabla^\mu h_{\rho\sigma}
+h_{\rho\sigma}\nabla^\mu\Phi^{\rho\sigma}
-\Phi^{\mu\nu}\nabla_\nu h
+h\nabla_\nu\Phi^{\mu\nu}
-2h^{\rho\mu}\nabla_\rho X\nn \\
&&+4X g^{\rho[\mu}\nabla^{\sigma]}h_{\rho\sigma}
+2h\nabla^\mu X +4\delta \phi G^{\mu\nu} \nabla_\nu \phi
\big]
\, , \label{ThetaofWadS}
\eea
while the off-shell Noether potentials are
\bea
K_{(s)}^{\mu\nu}&=&2\lambda\nabla^{[\mu}\zeta^{\nu]}
+2\beta\big(2\zeta^{[\mu}\nabla^{\nu]}X+X\nabla^{[\mu}\zeta^{\nu]}
+\zeta^{[\mu}\nabla_\sigma\Phi^{\nu]\sigma}
-\zeta_\sigma\nabla^{[\mu}\Phi^{\nu]\sigma}\nn \\
&&+\Phi^{\sigma[\mu}\nabla_\sigma\zeta^{\nu]}
\big)
\, . \label{OffSheNoPofWAdS}
\eea
Note that $K_{(s)}^{\mu\nu}$ in the above equation is equivalent to the on-shell Noether
potential obtained through Wald's covariant phase space approach in
\cite{FengLLP}, where the Noether potential was adopted to calculate the
mass of static black holes.
For the variation of the Noether potential (\ref{OffSheNoPofWAdS}) see
Eq. (\ref{VarofNoePet}) in the appendix \ref{appB}. Substituting the
expressions for $K_{(s)}^{\mu\nu}$ and $\Theta_{(s)}^\mu$ into Eq. (\ref{QdefineAn})
in the condition that $\zeta^\mu$ is a Killing vector, one can further
propose a formulation of the conserved charge in the scalar-tensor theory
described by the Lagrangian (\ref{WAdSLagran}).

Till now, it has been extensively studied to seek solutions of the
Lagrangian (\ref{WAdSLagran}), for instance, in the contexts of static
solutions with various asymptotical structures
\cite{FengLLP,CharmT,MomHGRF,KobaTana,Minami,AnabCO,BabiCh,Rinaldi,BabiCH,Cistern,KolyKPS},
rotating black holes in three dimensions \cite{BTZBraH,Giribet}, and slowly
rotating black holes in four dimensions \cite{MaSilMiBe,CisCDS}. Thermodynamics
of this theory was investigated in \cite{HuangGLY}. The formulation
(\ref{QdefineAn}) provides another avenue to obtain conserved charges of these
black holes. As an explicit example, we now pay attention to computing mass and
angular momentum of the BTZ-type black hole with a single scalar degree of freedom
in \cite{BTZBraH}, where the authors only utilized the method of Euclidean action
to calculate the mass of the black hole in the static case. The corresponding Lagrangian
of the black hole solution is the one in Eq. (\ref{WAdSLagran}) with $\lambda=1$. The line
element and the scalar field that is only dependent on the radial coordinate
take the forms
\bea
ds^2&=& -f(r)dt^2+\frac{dr^2}{f(r)}+r^2\Big(d\varphi-\frac{a}{2r^2}dt\Big)^2
 \, , \nn \\
\Big(\frac{d\phi}{dr}\Big)^2&=&-\frac{\ell^2\Lambda+1}{\beta f(r)}
\, , \label{BTZscaf}
\eea
where
\be
f(r)=\frac{r^2}{\ell^2}-m+\frac{a^2}{4r^2} \, , \quad
\ell^2=\frac{\beta}{\eta} \, ,
\ee
and the constants $(m,a)$ correspond to the mass and angular momentum
respectively.

The mass $M$ of the BTZ-type black hole (\ref{BTZscaf}) can be treated as a Noether
charge with respect to time translational symmetry reflected by the Killing vector
$\xi_{(t)}^\mu=(-1,0,0)$. On the other hand, the perturbations of the fields are achieved
by letting the parameters fluctuate as
$m\rightarrow m+dm$ and $a\rightarrow a+da$.
Under such conditions, the $(t,r)$ component of the ADT potential corresponding to the
Killing vector $\xi_{(t)}^\mu$ is read off as
\be
Q_{ADT}^{tr}=\frac{1}{2}\frac{1}{\sqrt{-g}}
\delta\big(\sqrt{-g}K_{(s)}^{tr}(\xi_{(t)})\big)
-\xi_{(t)}^{[t}\Theta_{(s)}^{r]}
=\frac{(1-\Lambda\ell^2)}{4r}dm \, . \label{trcomQBTZ}
\ee
Substituting Eq. (\ref{trcomQBTZ}) into the formulation (\ref{QdefineAn}), we have
\be
M=\frac{1}{16}(1-\Lambda\ell^2)m \, . \label{MassBTZ}
\ee
The angular momentum $J$ of the black hole can be obtained in a similar manner
as we perform to compute the mass when the Killing vector is chosen as
$\xi_{(\varphi)}^\mu=\delta^\mu_\varphi$. It is presented by
\be
J=\frac{1}{16}(1-\Lambda\ell^2)a \, . \label{AngumoBTZ}
\ee
Both the mass $M$ and the angular momentum $J$ satisfy the first law of thermodynamics.
In particular, if $\ell^2=-\Lambda^{-1}$, the scalar field $\phi$ vanishes.
$M$ and $J$ reduce to the mass and angular momentum of the conventional BTZ
black hole, respectively. In the work \cite{BTZBraH}, the authors also constructed
solutions of black holes with a planar horizon in arbitrary dimensions. Making use
of the formulation (\ref{QdefineAn}) to compute the mass of these black holes, we
get their mass that is consistent with the one derived via the method of Euclidean
action.

Next, we calculate the mass of the four-dimensional charged spherically symmetric
black hole with an asymptotically locally AdS structure in \cite{Cistern}.
The Lagrangian associated with this black hole is
$L_{(s)}(\lambda=1,\eta\rightarrow\alpha/2,\beta\rightarrow\eta/2)+L_{em}$,
where $L_{em}=-\sqrt{-g}F_{\mu\nu}F^{\mu\nu}/4$. The black hole solution is
given by \footnote{We have set $\kappa=1$ in comparison with the solution in
\cite{Cistern}. Therein the $t$ component of the gauge field $A_0(r)$ has
several typos.}
\bea
ds^2&=&-F(r)dt^2+G(r)dr^2+r^2\big(d\theta^2+\sin^2\theta d\varphi^2\big)
 \, , \quad
 A=A_t(r) dt \, , \nn \\
\Big(\frac{d\phi}{dr}\Big)^2&=&-\frac{1}{32}\frac{\alpha^2
[4(\alpha+\Lambda\eta)r^4+\eta q^2]
[4(\alpha-\Lambda\eta)r^4+8\eta r^2-\eta q^2]^2}{r^6\eta(\alpha-\Lambda\eta)^2
(\alpha r^2+\eta)^3F(r)}
\, .\label{charged4DBH}
\eea
In the above equation,
\bea
F(r)&=&\frac{r^2}{l^2}+\frac{\sqrt{\eta\alpha}}{\alpha}
\frac{[4\eta(\alpha+\Lambda\eta)+\alpha^2 q^2]^2}{16\eta^2(\alpha-\Lambda\eta)^2}
\frac{\arctan(\sqrt{\eta\alpha}r/\eta)}{r}-\frac{m}{r}
+\frac{3\alpha+\Lambda\eta}{\alpha-\Lambda\eta}  \nn \\
&&+\frac{\alpha^2q^2}{48\eta(\alpha-\Lambda\eta)^2}
\frac{3(\alpha q^2+16\eta)r^2-q^2\eta}{r^4} \, , \qquad
l^2=\frac{3\eta}{\alpha}
\, , \nn\\
G(r)&=&\frac{1}{16}
\frac{\alpha^2[4(\alpha-\Lambda\eta)r^4+8\eta r^2-\eta q^2]^2}{
r^4(\alpha-\Lambda\eta)^2(\alpha r^2+\eta)^2F(r)}   \, ,
\eea
and the $t$ component of the U(1) gauge field
\bea
A_t(r)&=&
\frac{q\sqrt{\eta\alpha}[4\eta(\alpha+\Lambda\eta)+\alpha^2q^2]}{4\eta^{2}
(\alpha-\Lambda\eta)}
\arctan\Big(\frac{\sqrt{\eta\alpha}}{\eta}r\Big)
+\frac{\alpha(8\eta+\alpha q^2)}{4\eta(\alpha-\Lambda\eta)}\frac{q}{r}\nn \\
&&-\frac{\alpha}{12(\alpha-\Lambda\eta)}\frac{q^3}{r^3} \, .
\eea
The constants $(m,q)$ denote the mass and electric charge
respectively and the parameter $\Lambda$ is assumed to satisfy that $\Lambda<0$.
When $q=0$, the black hole (\ref{charged4DBH}) reduces to the neutral one
in \cite{AnabCO}. To get the mass of the black hole, the infinitesimal
variation of the  fields is determined by letting the constants $(m,q)$
change as $m\rightarrow m+dm$ and $q\rightarrow q+dq$, and the Killing
vector $\xi_{(t)}^\mu=-\delta^\mu_t$. By using Eq. (\ref{ADTpotenQ}),
one can get the ADT potential related to the gravitational field and the
scalar field. Besides, since the theory includes gauge field $A_\mu$, one
has to consider the contribution to the potential from the Lagrangian
$L_{em}$, which is read off as \cite{CiteHJPY}
\bea
Q_{em}^{\mu\nu}&=&\frac{1}{4}\xi^\sigma A_\sigma \big(hF^{\mu\nu}
+4h^{\rho[\mu}F^{\nu]}_{~~\rho}
+4g^{\alpha[\mu}g^{\nu]\beta}\partial_\alpha\delta A_\beta\big)
+\frac{1}{2}F^{\mu\nu}\xi^\sigma \delta A_\sigma \nn \\
&&+\xi^{[\mu}F^{\nu]\sigma}\delta A_\sigma \, . \label{Qemmunu}
\eea
Therefore, the ADT potential corresponding to the Lagrangian
$L_{(s)}+L_{em}$ is
$Q_{ADT}^{\mu\nu}\rightarrow Q_{total}^{\mu\nu}=Q_{ADT}^{\mu\nu}+Q_{em}^{\mu\nu}$.
The $(t,r)$ component is
\be
\sqrt{-g}Q_{total}^{tr}
=\frac{3-\Lambda l^2}{6}\sin\theta d(m) \, , \label{Qtr4DBH}
\ee
whose integration yields the mass
\be
\mathcal{M}=\frac{3-\Lambda l^2}{12}m \, .\label{Massof4DBH}
\ee
When $q=0$ and $\alpha=-\eta\Lambda$, the black hole (\ref{charged4DBH})
reduces to the well-known four-dimensional Schwarzschild-AdS black hole.
In such a case, $\mathcal{M}=m/2$ is just the mass of the
Schwarzschild-AdS black hole. In the work \cite{AnabCO}, the authors
also computed the mass of the neutral spherically symmetric black hole
through the method of Euclidean action, which is different from the
mass $\mathcal{M}$ here and does not recover the mass of the
Schwarzschild-AdS black hole.

Finally, we have applied the method in the present work to calculate the
conserved charges of Warped-AdS$_3$ black holes with a scalar field in
\cite{Giribet}. Unfortunately, both the mass and angular momentum are
zero. This maybe arise from the fact that the formulation (\ref{QdefineAn})
for the conserved charge is covariant and the warped-AdS$_3$ black
hole is locally equivalent to the warped-AdS$_3$ space, while the mass
and angular momentum of the latter vanish.
In order to get sensible results, we shall take into account this
point in the future work.

%%%%%%%%%%%%%%%%%%%%%%%%%%%%%%%%%%%%%%%%%%%%%%%%%%%%
\section{Summary}\label{secfour}
%%%%%%%%%%%%%%%%%%%%%%%%%%%%%%%%%%%%%%%%%%%%%%%%%%%%

We obtain the off-shell Noether current (\ref{OffShNoeCurr}) and its
corresponding potential (\ref{OffSheNoPoten}) in the context of the full Horndeski
gravity theory described by the Lagrangian (\ref{LagranHorn}). To
achieve this, we first derive the surface terms and equations of
motion from the variation of the Lagrangian. By lifting the conventional
ADT potential to the off-shell level, we further give a proposal on the
formulation (\ref{QdefineAn}) of the conserved charge in terms of the
off-shell ADT potential, which is actually equivalent to the Noether
potential via the covariant phase space approach. Our derivation
provides a general and systematic method to compute the conserved charge
in the Horndeski theory. Because of the generality of the Horndeski theory,
it is feasible to extend the formulation (\ref{QdefineAn}) to various
well-known scalar-tensor theories with a single scalar degree of freedom.

As an application of the general formalism, we derive the off-shell Noether
potential of a specific subclass of the full Horndeski gravity theory depicted
by the Lagrangian (\ref{WAdSLagran}), namely, the scalar-tensor theory with
non-minimal coupling between derivatives of a scalar field and the
Einstein tensor. In terms of the off-shell Noether potential and the surface
term of this special theory, we first explicitly compute both the mass and angular
momentum of the BTZ-type black hole (\ref{BTZscaf}), as well as the mass of
the four-dimensional charged spherically symmetric black hole (\ref{charged4DBH}).

%%%%%%%%%%%%%%%%%%%%%%%%%%%%%%%%%%%%%%%%%%%%%%%%%%%%%%%%%%%
\appendix
\section{The variation of the terms $L_{(2,3)}$ } \label{appA}
%%%%%%%%%%%%%%%%%%%%%%%%%%%%%%%%%%%%%%%%%%%%%%%%%%%%%%%%%%

In this appendix, we shall derive the expressions of the field equations and
surface terms from the variation of the Lagrangian terms $L_{(2)}$ and
$L_{(3)}$. In what follows, note that we use the notations
$(\Psi_{\alpha\beta})^2=(\nabla_\alpha\nabla_\beta\phi)
(\nabla^\alpha\nabla^\beta\phi)$ and
$(\Psi_{\alpha\beta})^3=(\nabla_\alpha\nabla_\beta\phi)
(\nabla^\alpha\nabla_\gamma\phi)(\nabla^\gamma\nabla^\beta\phi)$.

The variation of the Lagrangian terms $L_{(2)}$ and $L_{(3)}$ with respect to both the
fields $g_{\mu\nu}$ and $\phi$ is presented as
\be
\delta L_{(j)}=\sqrt{-g}\big[\mathcal{T}^{(j)}_{\mu\nu}\delta g^{\mu\nu}
+\mathcal{E}_{(\phi)}^{(j)}\delta\phi
+\nabla_\mu\Theta_{(j)}^\mu(\delta g,\delta\phi)\big] \, , \quad
j=2,3 \, . \label{VarL2and3}
\ee
For the Lagrangian $L_{(2)}$, the expressions of the field equation
$\mathcal{T}^{(2)}_{\mu\nu}$ and $\mathcal{E}_{(\phi)}^{(2)}$ are given by
\bea
\mathcal{T}^{(2)}_{\mu\nu}&=&\frac{1}{2}g_{\mu\nu}G^{(2)}_{,X}(\Box\phi)^2
+g_{\mu\nu}\nabla_\sigma\big(G^{(2)}_{,X}\Box\phi\big)\nabla^\sigma \phi
-2\nabla_{(\mu}\big(G^{(2)}_{,X}\Box\phi\big)\nabla_{\nu)} \phi
+G^{(2)}G_{\mu\nu}\nn \\
&&+\frac{1}{2}G^{(2)}_{,XX}\big((\Psi_{\alpha\beta})^2-(\Box\phi)^2\big)\Phi_{\mu\nu}
+2\nabla^\sigma\big(G^{(2)}_{,X}\Psi_{\sigma(\mu}\big)\nabla_{\nu)} \phi
-\frac{1}{2}RG^{(2)}_{,X}\Phi_{\mu\nu}\nn \\
&&+\frac{1}{2}g_{\mu\nu}G^{(2)}_{,X}(\Psi_{\alpha\beta})^2
-\nabla_\sigma\big(G^{(2)}_{,X}\Psi_{\mu\nu}\nabla^\sigma\phi\big)
-\nabla_\mu\nabla_\nu G^{(2)}
+g_{\mu\nu}\Box G^{(2)}
\, , \label{ExpreT2} \\
\mathcal{E}_{(\phi)}^{(2)}&=&2\Box\big(G^{(2)}_{,X}\Box\phi\big)
-\nabla_\mu\big[G^{(2)}_{,XX}\big((\Psi_{\alpha\beta})^2-(\Box\phi)^2\big)\nabla^\mu\phi\big]
 +\nabla_\mu\big(G^{(2)}_{,X}R\nabla^\mu\phi\big)\nn \\
&&-G^{(2)}_{,X\phi}\big((\Psi_{\alpha\beta})^2-(\Box\phi)^2\big)
-2\nabla_\mu\nabla_\nu\big(G^{(2)}_{,X}\Psi^{\mu\nu}\big)
+G^{(2)}_{,\phi}R
\, , \label{ExprePhi2}
\eea
and the surface term $\Theta_{(2)}^\mu(\delta g,\delta\phi)$ takes the form
\bea
\Theta_{(2)}^\mu&=&G^{(2)}_{,X}\Box\phi(h\nabla^\mu\phi-2h^{\mu\nu}\nabla_\nu\phi)
+2G^{(2)}_{,X}\Box\phi\nabla^\mu\delta\phi
-2\nabla^\mu\big(G^{(2)}_{,X}\Box\phi\big)\delta\phi\nn \\
&&+G^{(2)}_{,XX}\big((\Psi_{\alpha\beta})^2-(\Box\phi)^2\big)\nabla^\mu\phi\delta\phi
+G^{(2)}_{,X}(2\Psi^{\mu\rho}\nabla^\sigma\phi
-\Psi^{\rho\sigma}\nabla^\mu\phi)h_{\rho\sigma}
\nn \\
&&-2G^{(2)}_{,X}\Psi^{\mu\nu}\nabla_\nu\delta\phi
+2\nabla_\nu\big(G^{(2)}_{,X}\Psi^{\mu\nu}\big)\delta\phi
-h^{\mu\nu}\nabla_\nu G^{(2)}+G^{(2)}\nabla_\nu h^{\mu\nu}\nn \\
&&+h\nabla^\mu G^{(2)}-G^{(2)}\nabla^\mu h
-G^{(2)}_{,X}R\nabla^\mu\phi\delta\phi \, . \label{Thetmu2}
\eea
For the Lagrangian $L_{(3)}$, the expression for the equation of motion
$\mathcal{T}^{(3)}_{\mu\nu}$ is presented by
\be
\mathcal{T}^{(3)}_{\mu\nu}=\mathcal{T}^{(31)}_{\mu\nu}+\mathcal{T}^{(32)}_{\mu\nu}
\, , \label{ExpreT3}
\ee
where
\bea
\mathcal{T}^{(31)}_{\mu\nu}&=&
-\frac{1}{2}G^{(3)}_{,XX}\big[(\Box\phi)^3-3\Box\phi(\Psi_{\alpha\beta})^2
+2(\Psi_{\alpha\beta})^3\big]\Phi_{\mu\nu}
+3\nabla_\sigma\big(G^{(3)}_{,X}\Psi_{\mu\rho}\Psi_\nu^{~\rho}\nabla^\sigma \phi\big)\nn \\
&&-\frac{1}{2}g_{\mu\nu}\big\{3\nabla_\sigma\big[G^{(3)}_{,X}\big((\Psi_{\alpha\beta})^2-(\Box\phi)^2\big)
\big]\nabla^\sigma \phi
+2G^{(3)}_{,X}\big((\Psi_{\alpha\beta})^3-(\Box\phi)^3\big) \big\}\nn \\
&&+3\nabla_{(\mu}\big[G^{(3)}_{,X}\big((\Psi_{\alpha\beta})^2-(\Box\phi)^2\big)\big]\nabla_{\nu)} \phi
+6\nabla^\sigma\big(G^{(3)}_{,X}\Box\phi\Psi_{\sigma(\mu}\big)\nabla_{\nu)} \phi \nn \\
&&-3\nabla_\sigma\big(G^{(3)}_{,X}\Box\phi\Psi_{\mu\nu}\nabla^\sigma \phi\big)
-6\nabla_\sigma\big(G^{(3)}_{,X}\Psi^{\sigma\rho}\Psi_{\rho(\mu}\big)\nabla_{\nu)}\phi
\, , \label{ExpreL3T31}
\eea
and the component $\mathcal{T}^{(32)}_{\mu\nu}$, which is the contribution from the second
term $-6\sqrt{-g}G^{(3)}G_{\mu\nu}\Psi^{\mu\nu}$ of $L_{(3)}$ in Eq. (\ref{ComLagranHorn}),
is read off as
\bea
\mathcal{T}^{(32)}_{\mu\nu}&=&
3G^{(3)}_{,X}G_{\rho\sigma}\Psi^{\rho\sigma}\Phi_{\mu\nu}
+3G^{(3)}\big(R\Psi_{\mu\nu}+R_{\mu\nu}\Box\phi-4R_{\rho(\mu}\Psi_{\nu)}^{~\rho}\big)
-3\Box\big(G^{(3)}\Psi_{\mu\nu}\big)\nn \\
&&+3g_{\mu\nu}\big[\Box\big(G^{(3)}\Box\phi\big)
-\nabla_\rho\nabla_\sigma\big(G^{(3)}\Psi^{\rho\sigma}\big)
+G^{(3)}G_{\rho\sigma}\Psi^{\rho\sigma}\big]\nn \\
&&-3\nabla_\mu\nabla_\nu\big(G^{(3)}\Box\phi\big)
+6\nabla^\rho\nabla_{(\mu}\big(\Psi_{\nu)\rho}G^{(3)}\big)\nn \\
&&+3\nabla^\sigma\big[G^{(3)}(2G_{\sigma(\mu}\nabla_{\nu)}\phi
-G_{\mu\nu}\nabla_\sigma\phi)\big]
\, . \label{ExpreL3T32}
\eea
The expression $\mathcal{E}_{(\phi)}^{(3)}$ of the equation of motion for the matter
field $\phi$ is given by
\bea
\mathcal{E}_{(\phi)}^{(3)}&=&
\nabla_\mu\big[G^{(3)}_{,XX}\big((\Box\phi)^3
-3\Box\phi(\Psi_{\alpha\beta})^2
+2(\Psi_{\alpha\beta})^3\big)\nabla^\mu \phi\big]
-6G^{(3)}_{,\phi}G^{\rho\sigma}\Psi_{\rho\sigma}\nn \\
&&-6\nabla_\mu\nabla_\nu\big(G^{(3)}_{,X}\Box\phi\Psi^{\mu\nu}\big)
+6\nabla_\mu\nabla^\nu\big(G^{(3)}_{,X}\Psi^{\mu\sigma}\Psi_{\nu\sigma}\big)
-6G^{\rho\sigma}\nabla_\rho\nabla_\sigma G^{(3)}\nn \\
&&+3\Box\big[G^{(3)}_{,XX}(\Box\phi)^2\big]
-3\Box\big[G^{(3)}_{,X}(\Psi_{\alpha\beta})^2\big]
-6\nabla_\mu\big(G^{(3)}_{,X}G^{\rho\sigma}\Psi_{\rho\sigma}\nabla^\mu \phi\big)\nn \\
&&+G^{(3)}_{,X\phi}\big((\Box\phi)^3
-3\Box\phi(\Psi_{\alpha\beta})^2
+2(\Psi_{\alpha\beta})^3\big)
 \, . \label{ExpreL3Phi}
\eea
The surface term $\Theta_{(3)}^\mu(\delta g,\delta\phi)$ is also split in two components, namely,
\be
\Theta_{(3)}^\mu(\delta g,\delta\phi)=\Theta_{(31)}^\mu(\delta g,\delta\phi)
+\Theta_{(32)}^\mu(\delta g,\delta\phi) \, , \label{ThetaL3}
\ee
where the component $\Theta_{(31)}^\mu$ coming from the contribution
of the variation of the first term in $L_{(3)}$ has the form
\bea
\Theta_{(31)}^\mu&=&
-\frac{3}{2}G^{(3)}_{,X}(\Psi_{\alpha\beta})^2
(h\nabla^\mu\phi-2h^{\mu\nu}\nabla_\nu\phi)
+3h_{\rho\sigma}G^{(3)}_{,X}\Box\phi(2\Psi^{\sigma\mu}\nabla^\rho\phi
-\Psi^{\sigma\rho}\nabla^\mu\phi)\nn \\
&&+\frac{3}{2}G^{(3)}_{,X}(\Box\phi)^2
(h\nabla^\mu\phi+2\nabla^\mu\delta\phi-2h^{\mu\nu}\nabla_\nu\phi)
-3\nabla^\mu\big[G^{(3)}_{,X}(\Box\phi)^2\big]\delta\phi \nn\\
&&-3G^{(3)}_{,X}(\Psi_{\alpha\beta})^2\nabla^\mu\delta\phi
+3\delta\phi\nabla^\mu \big[G^{(3)}_{,X}(\Psi_{\alpha\beta})^2\big]
-6G^{(3)}_{,X}\Box\phi\Psi^{\mu\nu}\nabla_\nu\delta\phi \nn \\
&&+6\delta\phi\nabla_\nu\big(G^{(3)}_{,X}\Box\phi\Psi^{\mu\nu}\big)
+6G^{(3)}_{,X}\Psi^{\mu\sigma}\Psi_{\nu\sigma}\nabla^\nu\delta\phi
-6\delta\phi\nabla^\nu\big(G^{(3)}_{,X}\Psi^{\mu\sigma}\Psi_{\nu\sigma}\big) \nn \\
&&-G^{(3)}_{,XX}\big((\Box\phi)^3
-3\Box\phi(\Psi_{\alpha\beta})^2
+2(\Psi_{\alpha\beta})^3\big)\nabla^\mu\phi\delta\phi \nn\\
&&-3G^{(3)}_{,X}\Psi^\nu_{~\sigma}(2\Psi^{\sigma\mu}\nabla^\rho\phi
-\Psi^{\sigma\rho}\nabla^\mu\phi)h_{\rho\nu} \, ,
\label{ThetaL31}
\eea

and the component $\Theta_{(32)}^\mu$, which is just the surface term from the
variation of the second term $-6\sqrt{-g}G^{(3)}G_{\mu\nu}\Psi^{\mu\nu}$
in $L_{(3)}$, is presented by
\bea
\Theta_{(32)}^\mu&=&
6h_{\rho\sigma}\nabla^\sigma\big(G^{(3)}\Psi^{\mu\rho}\big)
-6G^{(3)}\Psi_{\sigma\rho}\nabla^\sigma h^{\rho\mu}
+3G^{(3)}\Psi_{\rho\sigma}\nabla^\mu h^{\rho\sigma}\nn \\
&&-3h^{\rho\sigma}\nabla^\mu\big(G^{(3)}\Psi_{\rho\sigma}\big)
+3G^{(3)}\Psi^{\mu\nu}\nabla_\nu h
-3h\nabla_\nu\big(G^{(3)}\Psi^{\mu\nu}\big) \nn \\
&&+3G^{(3)}\Box\phi\nabla_\rho h^{\rho\mu}
-3h^{\rho\mu}\nabla_\rho\big(G^{(3)}\Box\phi\big)
-3G^{(3)}\Box\phi\nabla^\mu h\nn \\
&&+3h\nabla^\mu\big(G^{(3)}\Box\phi\big)
+6G^{(3)}h_{\rho\sigma}G^{\mu\rho}\nabla^\sigma\phi
-3G^{(3)}h_{\rho\sigma}G^{\rho\sigma}\nabla^\mu\phi \nn \\
&&+6\delta\phi G^{\mu\nu}\nabla_\nu G^{(3)}
-6G^{(3)}G^{\mu\nu}\nabla_\nu\delta\phi
+6\delta\phi G^{(3)}_{,X}G^{\rho\sigma}\Psi_{\rho\sigma}\nabla^\mu\phi
\, . \label{ThetaL3p32}
\eea

In the works \cite{KobaYY,GaoSteer,MaSilMiBe}, the equations of motion
for the Horndeski theory were also presented but the surface terms were
absent. To compare the field equations in this work with the ones in
\cite{GaoSteer}, one
can find that all the equations of motion for the matter field $\phi$ coincide
with each other and the relationship between the expressions of the equations of
motion with respect to the gravitational field $g_{\mu\nu}$
is $-2\mathcal{T}^{(i)}_{\mu\nu}=T^{(i)}_{\mu\nu}$, where $T^{(i)}_{\mu\nu}$
is the notation in \cite{GaoSteer}.

%%%%%%%%%%%%%%%%%%%%%%%%%%%%%%%%%%%%%%%%%%%%%%%%%%%%%%%%
\section{The variation for the off-shell Noether potentials $K_{(i)}^{\mu\nu}$
and $K_{(s)}^{\mu\nu}$} \label{appB}
%%%%%%%%%%%%%%%%%%%%%%%%%%%%%%%%%%%%%%%%%%%%%%%%%%%%%%%%

In the present appendix, we give a derivation of the variation for the off-shell
Noether potentials $K_{(i)}^{\mu\nu}$ and $K_{(s)}^{\mu\nu}$ under the condition
that $\delta\zeta^\mu=0$. Varying the off-shell Noether potentials
$K_{(1)}^{\mu\nu}$ and $K_{(2)}^{\mu\nu}$ in Eq. (\ref{OffShePote012}), we have
\bea
\delta K_{(1)}^{\mu\nu}&=&2\delta G^{(1)}\zeta^{[\mu}\nabla^{\nu]}\phi
+2G^{(1)}\big(\zeta^{[\mu}\nabla^{\nu]}\delta\phi
-\zeta^{[\mu}h^{\nu]\sigma}\nabla_\sigma\phi\big)\, , \nn \\
\delta K_{(2)}^{\mu\nu}&=&4\delta G_{,X}^{(2)}
\big(\Box\phi\zeta^{[\mu}\nabla^{\nu]}\phi
-\zeta_\sigma\Psi^{\sigma[\mu}\nabla^{\nu]}\phi\big)
+4G_{,X}^{(2)}\big[\delta(\Box\phi)\zeta^{[\mu}\nabla^{\nu]}\phi\nn \\
&&+\Box\phi\big(\zeta^{[\mu}\nabla^{\nu]}\delta\phi
-\zeta^{[\mu}h^{\nu]\sigma}\nabla_\sigma\phi\big)
-h_{\rho\sigma}\zeta^\rho\Psi^{\sigma[\mu}\nabla^{\nu]}\phi
-\zeta_\sigma\delta(\Psi^{\sigma[\mu}\nabla^{\nu]}\phi)\big]\nn \\
&&-4\zeta^{[\mu}h^{\nu]\sigma}\nabla_\sigma G^{(2)}
+4\zeta^{[\mu}\nabla^{\nu]}\delta G^{(2)}
+2G^{(2)}(\zeta_\sigma\nabla^{[\mu}h^{\nu]\sigma}
-h^{\sigma[\mu}\nabla_\sigma\zeta^{\nu]})\nn \\
&&+ 2\delta G^{(2)}\nabla^{[\mu}\zeta^{\nu]}
\, , \label{VaOffShePote012}
\eea
where
\bea
\delta X&=&\frac{1}{2}h_{\rho\sigma}\Phi^{\rho\sigma}
-(\nabla^{\sigma}\phi)(\nabla_{\sigma}\delta\phi) \, , \nn \\
\delta\Psi_{\mu\nu} &=& \nabla_\mu\nabla_\nu \delta\phi
-\frac{1}{2}(2\nabla_{(\mu} h_{\nu)\lambda}
-\nabla_\lambda h_{\mu\nu})\nabla^\lambda\phi \, , \nn \\
\delta G^{(i)}&=&G_{,\phi}^{(i)}\delta\phi+G_{,X}^{(i)}\delta X \, ,\quad
\delta G_{,X}^{(i)}=G_{,X\phi}^{(i)}\delta\phi+G_{,XX}^{(i)}\delta X \, , \nn \\
\delta\Psi^{\mu\nu}&=&-2h^{\sigma(\mu}\Psi^{\nu)}_{~\sigma}
+g^{\mu\rho}g^{\nu\sigma}\delta\Psi_{\rho\sigma} \, , \quad
\delta\Box\phi=h_{\rho\sigma}\Psi^{\rho\sigma}
+g_{\rho\sigma}\delta\Psi^{\rho\sigma}  \, .
\eea
and
\be
\delta(\Psi^{\sigma[\mu}\nabla^{\nu]}\phi)=
\delta\Psi^{\sigma[\mu}\nabla^{\nu]}\phi
-\Psi^{\sigma[\mu}h^{\nu]\rho}\nabla_\rho\phi
+\Psi^{\sigma[\mu}\nabla^{\nu]}\delta\phi \, .
\ee
The perturbation of $K_{(3)}^{\mu\nu}$ in Eq. (\ref{OffShelPoten3}) takes the form
\be
\delta K_{(3)}^{\mu\nu}=\delta K_{(31)}^{\mu\nu}+\delta K_{(32)}^{\mu\nu}
\, , \label{VaOffShPoten3}
\ee
where
\bea
\delta K_{(31)}^{\mu\nu}&=&6\delta G^{(3)}_{,X}\big[\big((\Box\phi)^2
-\Psi_{\alpha\beta}\Psi^{\alpha\beta}\big)\zeta^{[\mu}\nabla^{\nu]}\phi
+2\Psi^{\sigma[\mu}\nabla^{\nu]}\phi\big(\zeta^\rho\Psi_{\rho\sigma}
-\Box\phi\zeta_\sigma\big)\big]\nn \\
&&+6G^{(3)}_{,X}\big[(2\Box\phi\delta\Box\phi
-\delta\Psi_{\alpha\beta}\Psi^{\alpha\beta}
-\Psi_{\alpha\beta}\delta\Psi^{\alpha\beta})\zeta^{[\mu}\nabla^{\nu]}\phi
+\big((\Box\phi)^2\nn \\
&&-\Psi_{\alpha\beta}\Psi^{\alpha\beta}\big)
\big(\zeta^{[\mu}\nabla^{\nu]}\delta\phi
-\zeta^{[\mu}h^{\nu]\sigma}\nabla_\sigma\phi\big)
+2\delta(\Psi^{\sigma[\mu}\nabla^{\nu]}\phi)\big(\zeta^\rho\Psi_{\rho\sigma}\nn \\
&&-\Box\phi\zeta_{\sigma}\big)
+2\big(\zeta^\rho\delta\Psi_{\rho\sigma}-\zeta_\sigma\delta\Box\phi
-h_{\sigma\lambda}\zeta^\lambda\Box\phi\big)\Psi^{\sigma[\mu}\nabla^{\nu]}\phi\big]\,
\label{VaOffShK31}
\eea
and
\bea
\delta K_{(32)}^{\mu\nu}&=&-6\big[2\zeta^{[\mu}\delta Y_\sigma^{~\nu]\sigma}
-2h_{\rho\sigma}\zeta^\rho Y^{[\mu\nu]\sigma}
+2\zeta_\sigma h^{\rho[\mu}Y_\rho^{~\nu]\sigma}
-2\zeta_\sigma g^{\rho[\mu}\delta Y_\rho^{~\nu]\sigma} \nn \\
&&-2h_{\rho\sigma}\zeta^{[\mu}Y^{\nu]\rho\sigma}
+2g_{\rho\sigma}\zeta^{[\mu}h^{\nu]\lambda}Y_\lambda^{~\rho\sigma}
-2g_{\rho\sigma}\zeta^{[\mu}g^{\nu]\lambda}\delta Y_\lambda^{~\rho\sigma}\nn \\
&&+2\big(G^{(3)}h_{\rho\sigma}\zeta^\rho
+\delta G^{(3)}\zeta_\sigma\big)G^{\sigma[\mu}\nabla^{\nu]}\phi
+2G^{(3)}\zeta_\sigma\big(\delta G^{\sigma[\mu}\nabla^{\nu]}\phi\nn \\
&&-G^{\sigma[\mu}h^{\nu]\rho}\nabla_\rho\phi
+G^{\sigma[\mu}\nabla^{\nu]}\delta\phi\big)
+2\delta G^{(3)}\Psi^{\sigma[\mu}\nabla_\sigma\zeta^{\nu]}
+G^{(3)}\zeta^\rho\big(\nn \\
&&\Psi^{\sigma[\mu}\nabla_\sigma h^{\nu]}_{~\rho}
+\Psi_\sigma^{~[\mu}\nabla_\rho h^{\nu]\sigma}
-\Psi^{\sigma[\mu}\nabla^{\nu]} h_{\rho\sigma}\big)
+2G^{(3)}\delta\Psi^{\sigma[\mu}\nabla_\sigma\zeta^{\nu]}\nn \\
&&-\delta\big(G^{(3)}\Box\phi\big)\nabla^{[\mu}\zeta^{\nu]}
-G^{(3)}\Box\phi\big(\zeta_\sigma\nabla^{[\mu}h^{\nu]\sigma}
- h^{\sigma[\mu}\nabla_\sigma\zeta^{\nu]}\big)\big]
\, . \label{VaOffShPK32}
\eea
In the above equation,
\bea
Y_\rho^{~\mu\nu}&=&\nabla_\rho\big(G^{(3)}\Psi^{\mu\nu}\big) \, , \nn \\
\delta Y_\rho^{~\mu\nu}&=&\nabla_\rho\big(\delta G^{(3)}\Psi^{\mu\nu}
+G^{(3)}\delta\Psi^{\mu\nu}\big)
+G^{(3)}g^{\alpha(\mu}\Psi^{\nu)\beta}(2\nabla_{(\rho} h_{\beta)\alpha}
-\nabla_\alpha h_{\rho\beta}) \, , \nn \\
\delta G^{\sigma\mu}&=&\frac{1}{2}\big[2\nabla_\lambda\nabla^{(\sigma}h^{\mu)\lambda}
-4h_\lambda^{~(\sigma}R^{\mu)\lambda}-\Box h^{\sigma\mu}
-\nabla^\sigma\nabla^\mu h+Rh^{\sigma\mu} \nn \\
&&-g^{\sigma\mu}(\nabla^\alpha\nabla^\beta h_{\alpha\beta}
-h^{\alpha\beta}R_{\alpha\beta}-\Box h)\big] \, .
\eea
Besides, the variation of the total off-shell Noether potential
$K^{\mu\nu}$ can be expressed as $\delta K^{\mu\nu}=
\sum_{i=1}^{i=3}\delta K_{(i)}^{\mu\nu}$.

Finally, the variation of the off-shell Noether potential $K_{(s)}^{\mu\nu}$ in Eq.
(\ref{OffSheNoPofWAdS}) is given by
\bea
\delta K_{(s)}^{\mu\nu}&=&
2(\lambda+\beta X)(\zeta_\sigma\nabla^{[\mu}h^{\nu]\sigma}
-h^{\sigma[\mu}\nabla_\sigma\zeta^{\nu]})
+2\beta(\delta X)\nabla^{[\mu}\zeta^{\nu]}
+\beta\big[4\zeta^{[\mu}\nabla^{\nu]}\delta X\nn \\
&&-4\zeta^{[\mu}h^{\nu]\sigma}\nabla_\sigma X
+2\zeta^{[\mu}\delta(\nabla_\sigma\Phi^{\nu]\sigma})
-2\zeta_\sigma g^{\rho[\mu}\delta(\nabla_\rho\Phi^{\nu]\sigma})
-2h_{\rho\sigma}\zeta^\rho\nabla^{[\mu}\Phi^{\nu]\sigma}\nn \\
&&+2\zeta_\sigma h^{\rho[\mu}\nabla_\rho\Phi^{\nu]\sigma}
+2(\delta\Phi^{\sigma[\mu})\nabla_\sigma\zeta^{\nu]}
+\zeta_\rho\Phi^{\sigma[\mu}\nabla_\sigma h^{\nu]\rho}
+\zeta^\rho\Phi_\sigma^{~[\mu}\nabla_\rho h^{\nu]\sigma}\nn \\
&&-\zeta^\rho\Phi^{\sigma[\mu}\nabla^{\nu]} h_{\rho\sigma}
\big]\, , \label{VarofNoePet}
\eea
where
\bea
\delta\Phi^{\mu\nu}&=&-2h^{\sigma(\mu}\Phi^{\nu)}_{~\sigma}
+2(\nabla^{(\mu}\phi)(\nabla^{\nu)}\delta\phi) \, ,
\nn \\
\delta(\nabla_\rho\Phi^{\nu\sigma})&=&\nabla_\rho\delta\Phi^{\nu\sigma}
+g^{\alpha(\nu}\Phi^{\sigma)\beta}(2\nabla_{(\rho} h_{\beta)\alpha}
-\nabla_\alpha h_{\rho\beta}) \, .
\eea

\section*{Acknowledgments}

This work was supported by the Natural Science Foundation of China under Grant
Nos. 11275157 and 11505036. It was also partially supported by the Doctoral
Research Fund of Guizhou Normal University in 2014 and
Guizhou province science and technology innovation talent
team [Grant No. (2015)4015].

%\paragraph{Note added.}

\end{document}